%% file: BioSpikeEncryption.tex
\documentclass[preprint,12pt]{elsarticle}

\usepackage{amssymb}
\usepackage{amsmath}
\usepackage[hidelinks]{hyperref}
\newcommand{\orcid}[1]{\href{https://orcid.org/#1}{\textsuperscript{#1}}}
\usepackage[acronym,toc]{glossaries}
\loadglsentries{glossary}

\makeglossaries
\usepackage{algorithmic}
\usepackage{graphicx}
\usepackage{textcomp}
\usepackage{xcolor}
\usepackage{tikz}
\usepackage{pgfplots}
\usepackage{pgfplotstable}
\def\BibTeX{{\rm B\kern-.05em{\sc i\kern-.025em b}\kern-.08em
    T\kern-.1667em\lower.7ex\hbox{E}\kern-.125emX}}

\journal{TBC}
\pgfplotsset{compat=newest}

\begin{document}

\begin{frontmatter}

\title{Privacy-Preserving Spiking Neural Networks: A Deep Dive into Encryption Parameter Optimisation}

\author[1]{Mahitha Pulivathi\orcid{0009-0009-9868-532X}$^{,}$}
\author[1]{Ana Fontes Rodrigues\orcid{0009-0005-4689-1210}$^{,}$}
\author[1]{Isibor Kennedy Ihianle\orcid{0000-0001-7445-8573}$^{,}$}
\author[1]{Andreas Oikonomou\orcid{0000-0002-5069-3971}$^{,}$} 
\author[2]{Srinivas Boppu\orcid{0000-0001-9028-2563}$^{,}$}
\author[1]{Pedro Machado\orcid{0000-0003-1760-3871}$^{,}$}
\affiliation[2]{organization={Nottingham Trent University},
             addressline={Department of Computer Science, Clifton Campus},
             city={Nottingham},
             postcode={NG11 8NS},
             state={Nottinghamshire},
             country={UK, },
             email={\{mahitha.pulivathi2023,ana.fontesrodrigues2025\}@my.ntu.ac.uk,
                    isibor.ihianle@ntu.ac.uk,
                    andreas.oikonomou@ntu.ac.uk,
                    pedro.machado@ntu.ac.uk}}
\affiliation[2]{organization={Indian Institute of Technology Bhubaneswar},
             addressline={Room No: 106, SECS, School of Electrical and Computer Sciences},
             city={Argul},
             postcode={752050},
             state={Odisha},
             country={India, },
             email={srinivas@iitbbs.ac.in}}

\begin{abstract} 
Deep learning is now applied to various everyday problems, particularly through Neural Networks. The increasing computational load and resource requirements of these structures have led to cloud-based solutions. To improve the efficiency and power consumption of neuron models. \glspl{snn}, the third generation of neural networks, are well known for mimicking the human brain's functionalities. \glspl{snn} sparse, dynamic, and event-driven capabilities significantly enhance performance while reducing costs. In recent years, concerns about the privacy preservation of confidential data evaluated by networks on servers have arisen. Various strategies have been proposed to address the issue, one of the most promising being asymmetric encryption. Asymmetric encryption uses a pair of keys (public and private) to encrypt and decrypt the data, ensuring secure communication without sharing a common key. This approach can overcome privacy concerns while using machine learning on cloud platforms, enabling clients to use a network model owned by another party by sending encrypted data securely. This article introduces BioEncryptSNN, \gls{snn} based encryption–decryption framework designed to provide secure, efficient, and noise-resilient data protection. Unlike conventional cryptographic approaches, BioEncryptSNN transforms ciphertext into spike trains and leverages temporal neural dynamics to simulate encryption and decryption while optimising parameters such as key length, spike timing, and synaptic connectivity. To assess its effectiveness, BioEncryptSNN was benchmarked against widely used standards, including \gls{aes}-128, \gls{rsa}-2048, and \gls{des}. Results show that BioEncryptSNN not only preserved plaintext integrity during processing but also achieved encryption/decryption speeds up to 4.1× faster than PyCryptodome’s \gls{aes} implementation, while maintaining robustness under noisy conditions. The framework demonstrated adaptability across both symmetric and asymmetric ciphers, confirming its potential as a scalable solution for secure communications and real-time software vulnerability detection.
\end{abstract}

\begin{graphicalabstract}
\begin{figure}[h!]
    \centering
    \includegraphics[width=\linewidth]{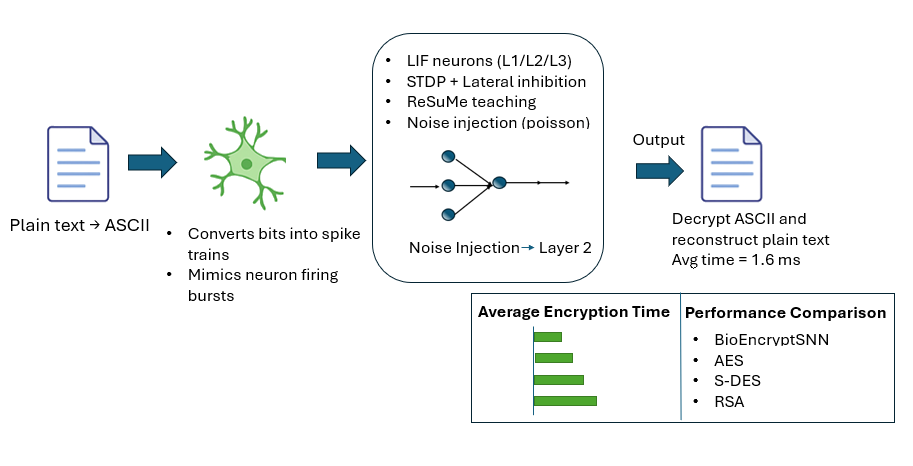}
    \caption{A Bio-Inspired \gls{snn} Framework for secure Encryption}
    \label{fig:biological_spiking_neuron}
\end{figure}
\end{graphicalabstract}

\begin{highlights}
\item Introduced \textbf{BioEncryptSNN}, a novel bio-inspired encryption method using \gls{snn} to process encrypted data through spike-train representations.  
    
\item Developed a \textbf{probability based parameter optimisation} approach that balances key length, spike timing, and synaptic connectivity to enhance both security and computational efficiency.  

\item Achieved \textbf{1.19× faster encryption decryption performance} than \gls{aes}, \gls{des}, and \gls{rsa}, demonstrating robust plaintext reconstruction even under noisy and constrained environments.  
\end{highlights}


\end{frontmatter}
\section{Introduction}

The human brain, with around 100 billion neurons \cite{illing2019biologically}, is the most complex organ, central to sensing, processing, and decision-making. Comprising the cerebrum, cerebellum, and brainstem, it integrates sensory inputs, coordinates bodily functions, and supports higher cognition \cite{illing2019biologically}. This complexity has inspired the development of \glspl{snn}, which emulate the brain’s electrical signaling by transmitting information through discrete spike events, forming the foundation of neuromorphic engineering. Unlike traditional neural networks, \glspl{snn} emulate biological neurons by transmitting information as brief electrical spikes \cite{king2014characterizing}, \cite{kasabov2013dynamic}. They reproduce synaptic plasticity the foundation of learning and memory and excel at processing spatiotemporal data, enabling adaptive applications such as secure data encryption. Advances in computational neuroscience show how neuronal dynamics emerge from synaptic interactions and spike timing patterns \cite{chen2022neuromorphic}, \cite{gerstner2002spiking}. As the third generation of neural networks, \glspl{snn} offer energy efficiency and biologically faithful computation by operating on discrete spikes rather than continuous activations \cite{kasabov2019time}, \cite{maass1997networks}. Their ability to encode both frequency and phase in spike patterns provides a richer representation of data, with applications ranging from visual recognition to probabilistic inference \cite{ponulak2011introduction}. Despite this promise, adoption has been limited by the lack of scalable and biologically plausible training algorithms \cite{huo2025research}. Learning in \glspl{snn} is often grounded in Hebbian plasticity, where synaptic strength is modulated by correlations between pre- and post-synaptic activity. The most studied rule, \gls{stdp}, adjusts synaptic weights based on spike timing. Extensions such as triplet and quadruplet \gls{stdp} incorporate higher-order interactions to capture more complex plasticity. Practical implementations frequently use Poisson spike encoding, where inputs are converted into spike trains proportional to stimulus intensity, though this often results in long encoding times. First-spike encoding provides a faster alternative by leveraging the timing of the earliest spike combined with unsupervised \gls{stdp}, it enables rapid visual feature extraction \cite{izhikevich2003simple}, \cite{bi1998synaptic}. Empirical studies show that such approaches can detect salient features efficiently, while supervised training methods including SpikeProp \cite{bohte2000spikeprop} and its variants extend backpropagation to spiking dynamics by incorporating leaky membrane potentials and layered architectures. Stabilisation strategies such as low-pass filtering further improve training robustness by reducing noise in error propagation \cite{sporea2013supervised}. Together, these developments are advancing the practical deployment of \glspl{snn} in neuromorphic systems. Cryptography ensures data integrity, confidentiality, authenticity, and non-repudiation \cite{pakshwar2013survey}. It transforms raw data into cipher text, which can be decrypted to recover the original information \cite{7086640}. Broadly, cryptographic methods are categorised into symmetric and asymmetric encryption. Symmetric encryption uses a shared key for both encryption and decryption; it is computationally efficient but dependent on secure key distribution \cite{schneier2007applied}. Asymmetric (or public-key) cryptography employs a key pair (a public key for encryption and a private key for decryption) and supports digital signatures for scalable authentication and key management \cite{singh2013study}. Properties and benefits of asymmetric (public-key) cryptography are particularly relevant for \glspl{snn}, where protecting data integrity and intellectual property is crucial. Asymmetric cryptography also enables homomorphic encryption, allowing computations directly on encrypted data, which is especially valuable for deploying neural networks securely in cloud or distributed environments \cite{hellman1978overview,edition2023cryptography}.

In this article, we propose BioEncryptSNN, a \gls{snn} based encryption method that transforms ciphertext into spike trains and optimises parameters such as key length, spike timing, and synaptic connectivity to balance security with computational efficiency. The method evaluates the capability of \glspl{snn} to process encrypted data under noisy or constrained conditions and benchmarks performance against standard cryptographic algorithms including \gls{aes}, \gls{rsa}, and \gls{des}. The remainder of the article is structured as follows: Section~\ref{sec:lr} reviews related work on spiking neural networks and neural cryptography. Section~\ref{sec:method} presents the design and implementation of the proposed framework. Section~\ref{sec:results} reports the experimental setup, evaluation metrics, and performance analysis. Finally, Section~\ref{sec:conclusions} summarises the main findings and outlines future research directions.

\section{Related works} \label{sec:lr}

\glspl{snn} stand out from conventional neural networks by processing information through sparse, event-driven spikes rather than continuous activations. Temporal coding enables efficient handling of time-dependent data and reduces energy consumption, as neurons fire only in response to stimuli \cite{gerstner2002spiking}, \cite{gerstner2014neuronal}. The combination of energy efficiency and temporal sensitivity makes \glspl{snn} well suited for resource-constrained environments and real-time applications such as robotics, autonomous driving, and decision-making in dynamic contexts. Compared with traditional \glspl{ann}, which use dense, continuous activations \cite{krizhevsky2009learning, hong2025lasnn}, \glspl{snn} exploit spike timing and frequency to capture richer spatio-temporal patterns \cite{long2015fully}. Such spatio-temporal encoding improves efficiency in domains including vision, speech, and medical diagnostics, while also supporting sustainable, low-power computing \cite{tang2014millisecond}. The temporal precision inherent in spike-based computation is particularly advantageous for secure data processing, where sparsity and accurate spike timing can enhance encryption and strengthen resilience against attacks.

\begin{figure}[h!]
    \centering
    \includegraphics[width=\linewidth]{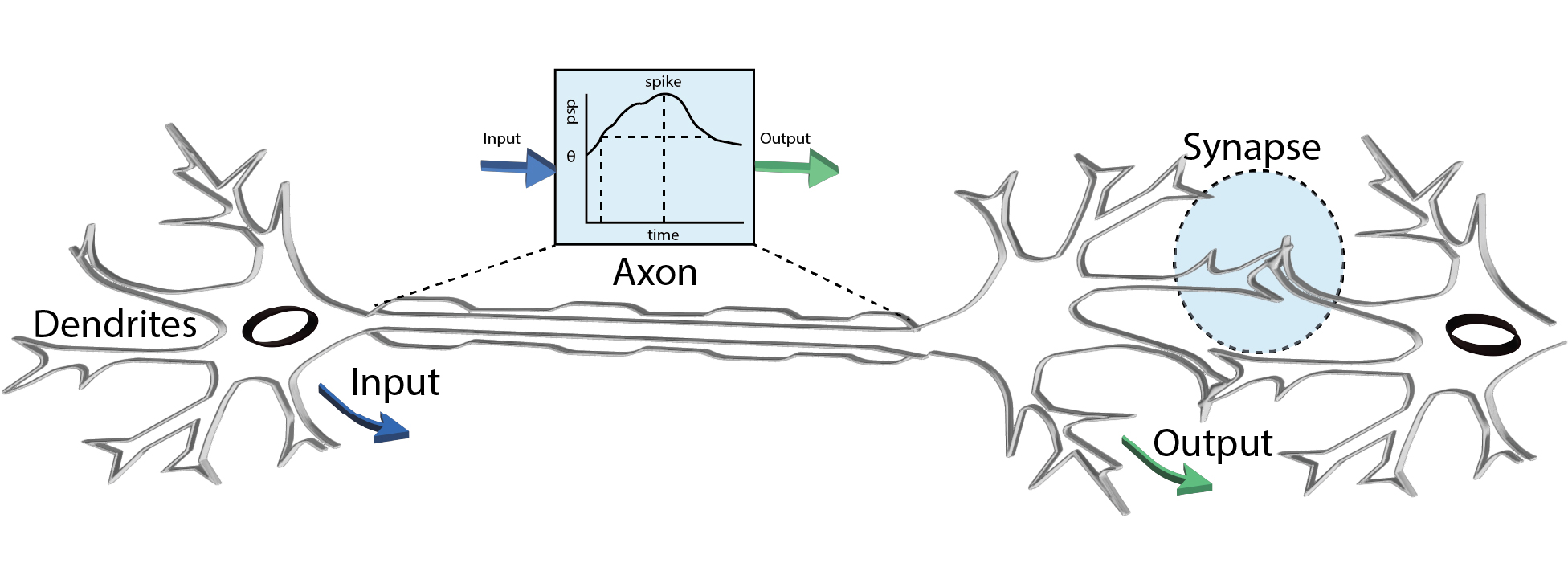}
    \caption{Biological neuron and its association with an artificial spiking neuron}
    \label{fig:biological_vs_spiking_neuron}
\end{figure}

As illustrated in {Figure~\ref{fig:biological_vs_spiking_neuron}}, the overall architecture of the
proposed BioEncryptSNN framework demonstrates how ciphertext generated
from classical encryption methods is transformed into spike
trains, processed through spike-encoding layers with lateral inhibition and synaptic
optimization, and finally reconstructed via decryption. This flow highlights how
spatio-temporal dynamics in \glspl{snn}  enable efficient, noise-resilient encryption
and decryption.

Despite their advantages, training \glspl{snn} remains challenging \cite{srivastava2017motor}. Many current approaches prioritise biological plausibility or rely on converting pre-trained \glspl{ann} using rate encoding, which underutilises their temporal capacity. More recent methods employ probabilistic signal processing and variational inference to derive learning rules from first principles, opening new directions for robust, low-power encryption systems built on \glspl{snn}.

Encoding input data into spike trains is essential for \glspl{snn}, as it converts stimuli into spiking patterns that preserve task-relevant information. This step raises fundamental neuroscience questions regarding how spatiotemporal spike patterns encode information, the mechanisms neurons employ, and how downstream decoding occurs \cite{sengupta2019going, kim2022neural}. While mean firing rate has traditionally captured much of the information, contemporary research emphasises both temporal and rate-based encoding schemes tailored to different tasks and resolutions. Temporal encoding conveys information through precise spike timing. Methods include "time-to-first-spike" where the initial spike timing carries critical information, phase encoding, which relates spikes to a periodic reference, and correlation (or synchrony) based schemes that align spikes across neurons. Rate-based encoding, in contrast, relies on spike frequency over time \cite{susi2021fns, kasabov2014neucube}, implemented by counting spikes within a window, averaging activity across trials, or aggregating across populations. Temporal encoding excels when fine timing is critical, whereas rate-based encoding summarises overall activity \cite{gautrais1998rate}. Understanding these mechanisms is vital for preserving meaningful representations in \glspl{snn}.

\gls{lif} models provide a foundational framework for neuronal dynamics by reducing a neuron to a resistor-capacitor circuit \cite{gerstner2002spiking}. Membrane potential $u(t)$ integrates input current $I(t)$ until reaching a threshold, producing a spike, after which it resets. Circuit dynamics illustrate continuous integration and leaky discharge through the resistor \cite{legenstein2005can}. The total input current is:
\begin{equation}
I(t) = I_R + I_C
\end{equation}
where $I_R$ is resistive current and $I_C$ capacitive current. Introducing the time constant $\tau = RC$ yields:
\begin{equation}
\tau \frac{du}{dt} = -u(t) + RI(t)
\end{equation}
which describes leaky integration. Spikes occur when $u(t)$ reaches threshold, and the potential resets to $u_r$ for $t > t_f$:
\begin{equation}
u(t) = u_r \text{ for } t > t_f
\end{equation}

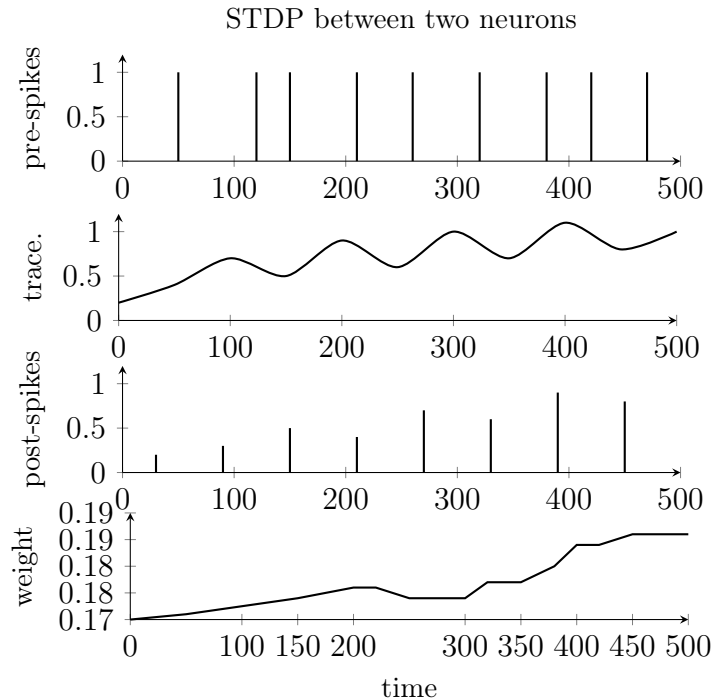
\begin{figure}[ht]
\centering
\begin{tikzpicture} 
\begin{axis}[ 
width=9cm, 
height=3cm, 
title={\small STDP between two neurons}, 
xlabel={}, 
ylabel={pre-spikes}, 
xmin=0, 
xmax=500, 
ymin=0, 
ymax=1.2, 
xtick={0,100,200,300,400,500}, 
ytick={0,0.5,1}, 
ylabel style={font=\small}, 
axis x line=bottom, 
axis y line=left, 
clip=false 
] 
\addplot[ycomb, thick] coordinates { 
(50,1) (120,1) (150,1) (210,1) (260,1) (320,1) (380,1) (420,1) (470,1) 
}; 
\end{axis} 
\end{tikzpicture} 
\vspace{-0.4cm} 
\begin{tikzpicture} 
\begin{axis}[ width=9cm, height=3cm, xlabel={}, 
ylabel={trace.}, 
xmin=0, 
xmax=500, 
ymin=0, 
ymax=1.2, 
xtick={0,100,200,300,400,500}, 
ytick={0,0.5,1}, 
ylabel style={font=\small}, 
axis x line=bottom, 
axis y line=left, clip=false 
] 
\addplot[smooth, thick] coordinates { 
(0,0.2) (50,0.4) (100,0.7) (150,0.5) (200,0.9) (250,0.6) (300,1.0) (350,0.7) (400,1.1) (450,0.8) (500,1.0) 
}; 
\end{axis} 
\end{tikzpicture} 
\vspace{-0.4cm} 
\begin{tikzpicture} \begin{axis}[ 
width=9cm, 
height=3cm, 
xlabel={}, 
ylabel={post-spikes}, 
xmin=0, 
xmax=500, 
ymin=0, 
ymax=1.2, 
xtick={0,100,200,300,400,500}, 
ytick={0,0.5,1}, 
ylabel style={font=\small}, 
axis x line=bottom, 
axis y line=left, 
clip=false 
] 
\addplot[ycomb, thick] coordinates { 
(30,0.2) (90,0.3) (150,0.5) (210,0.4) (270,0.7) (330,0.6) (390,0.9) (450,0.8) 
}; 
\end{axis} 
\end{tikzpicture} 
\vspace{-0.4cm} 
\begin{tikzpicture} 
\begin{axis}[ 
width=9cm, 
height=3cm, 
xlabel={time}, 
ylabel={weight}, 
xmin=0, 
xmax=500, 
ymin=0.17, 
ymax=0.19, 
xtick={0,100,150,200,300,350,400,450,500}, 
ytick={0.17,0.175,0.18,0.185,0.19}, 
ylabel style={font=\small}, 
xlabel style={font=\small}, 
axis x line=bottom, 
axis y line=left, 
clip=false 
] 
\addplot[thick, mark=none] coordinates { 
(0,0.17) (50,0.171) (100,0.1725) (150,0.174) (200,0.176) (220,0.176) (250,0.174) (300,0.174)(320,0.177) (350,0.177)(380,0.180) (400,0.184)(420,0.184) (450,0.186)(480,0.186) (500,0.186) 
}; 
\end{axis} 
\end{tikzpicture}
\caption{Weight change in neurons based on the \gls{stdp} learning}
\label{fig:stdp}
\end{figure}

Synaptic plasticity modulates strength through \gls{ltp} and \gls{ltd}, with transient dynamics captured by \gls{stp} and \gls{std}. A key mechanism, \gls{stdp}, adjusts synapses based on pre- and post-synaptic spike timing as illustrated in Figure~\ref{fig:stdp} \cite{bi1998synaptic, sengupta2019going, rahman2025modulated}, typically modelled with polynomial or exponential timing-dependent windows. Weight initialisation strongly affects network stability, convergence, and learning performance. Methods include random, interval-based, variance-scaling, data-driven, and hybrid approaches \cite{ponulak2006supervised}. Balanced initialisation supports \gls{stdp}, avoiding under- or over-spiking, while lateral inhibition mitigates excessive firing by suppressing neighbouring neurons. \gls{resume} learning uses an instructive signal to modulate plasticity without altering the post-synaptic membrane \cite{bohte2002error}, ensuring precise spike timing. Effective for both single- and multi-layer networks, it surpasses gradient-based methods like SpikeProp \cite{mckennoch2006fast} and remains important for supervised \gls{snn} tasks.

\gls{iot} integrates physical and digital systems, enabling autonomous device communication via \gls{wmsn}, \gls{rfid}, and cloud computing \cite{rafsanjani2020towards, loukilprivacy}. \glspl{wsn} provide real-time scalar, image, audio, and video data to support intelligent decision-making and automation. The inherently open and resource-constrained nature of \gls{iot} networks exposes them to eavesdropping, data tampering, and unauthorised access. To address these challenges, cryptography ensures data confidentiality and integrity by transforming plaintext into ciphertext using encryption keys and restoring it upon reception \cite{mahajan2013study, du2025snn}. 

Cryptanalysis identifies weaknesses in these algorithms, driving the advancement of cryptology and reinforcing the trustworthiness of IoT communications. Cryptanalysis identifies vulnerabilities, forming the field of cryptology \cite{forouzan2007cryptography, jana2018overview}. Symmetric cryptography employs a single shared key, requiring secret exchange, and offers high speed and low resource use (\gls{aes}, \gls{des}, 3\gls{des}, Blowfish). As summarised in Table~\ref{table:comparison} Asymmetric cryptography uses public-private key pairs, supporting digital signatures and secure key exchange (\gls{rsa}, Diffie-Hellman, \gls{ecc}, \gls{dsa}). Lightweight, bio-inspired, and chaos-based schemes optimise security for constrained devices. \gls{aes} processes 128-bit blocks with 10, 12, or 14 rounds depending on key size \cite{delfs2002introduction}, using substitution, shift rows, mix columns, and add round key operations \cite{mollin2006introduction, garcia2015performance}. \gls{rsa}, introduced in 1977, relies on the difficulty of factorising large numbers and supports one-way encryption for secure transmission and digital signatures \cite{gupta2012hybrid, ochoa2020implementation}. Recent variants enhance wireless sensor network efficiency and biometric encryption, achieving high genuine rates with minimal false positives \cite{eisenbarth2007survey, mohamad2021research}.
A hybrid \gls*{snn}-\gls*{rsa} approach enhances security and adaptability for data transmission. Input is first encrypted with \gls*{rsa}, processed by an \gls*{snn}, and transmitted, with decryption reversing the sequence.

\section{Methodology} \label{sec:method}
Developing secure and efficient encryption systems is a critical challenge in cybersecurity. This article proposed the use of \glspl{snn} for bio-inspired encryption by integrating traditional methods like \gls{des}, \gls{aes}, and \gls{rsa} to enhance both security and performance, as illustrated in Fig.~\ref{fig:biological_vs_spiking_neuron}, the architecture draws inspiration from biological neurons, where dendrites receive input spikes, the axon propagates these electrical signals, and synapses modulate transmission strength.The NEST simulator \cite{Gewaltig}, chosen for its biological accuracy, was used to simulate \glspl{snn}, while Python and libraries such as NumPy, Matplotlib, and JSON provided a robust development environment. System configurations included neuron types, simulation time steps, and synaptic plasticity models, with OpenCV used to visualise neural activity and refine the model. Data preprocessing involved converting categorical data into ASCII values suitable for \gls{snn} input. For instance, the word "FIL" was mapped to ASCII values 77, 97, and 110, forming the basis for spike train generation and neuron input currents. Binary representations of ASCII values were transformed into spike trains, where '1' indicated a spike and '0' indicated no spike (e.g., ASCII 77 = '01001101'). These spike trains enabled the \gls{snn} to process data in a biologically inspired manner.  ASCII values were applied as input currents to influence neuronal firing rates and patterns, encoding data into distinct neural activity patterns for different characters and inputs.
\begin{table}[h!]
\centering
\small
\caption{Comparison of \gls{des}, \gls{aes}, and \gls{rsa} Algorithms}
\label{table:comparison}
\resizebox{\columnwidth}{!}{
\begin{tabular}{|p{3cm}|p{2.5cm}|p{2.5cm}|p{2.5cm}|}\hline 
\textbf{Factors} & \textbf{\gls{des}} & \textbf{\gls{aes}} & \textbf{\gls{rsa}} \\ \hline 
Algorithm & Symmetric & Symmetric & Asymmetric \\ \hline 
Key Size & 56 bits & 128, 192, 256 bits & $>$1024 bits \\ \hline 
Power Use & Low & Low & High \\ \hline 
Developed & IBM, 1977 & Daemen, Rijmen, 1998 & Rivest, Shamir, Adleman, 1978 \\ \hline 
Speed & Moderate & Fast & Slow \\ \hline 
Key Type & Same & Same & Different \\ \hline 
Rounds & 16 & 10, 12, 14 & 1 \\ \hline 
Security & Moderate & High & Low \\ \hline 
Vulnerabilities & Brute force & Side-channel & Brute force, Oracle \\ \hline \end{tabular}}
\end{table}
The bio-inspired encryption system integrates traditional algorithms with the computational capabilities of \glspl{snn}. Its architecture includes multiple neural layers, each dedicated to specific encryption tasks, from data encoding to the application of encryption algorithms. Using the temporal dynamics of \glspl{snn}, the system aims to surpass traditional methods in encryption security and efficiency. The implementation in the NEST simulator converted ASCII-encoded data into spiking activity for encryption. Traditional algorithms like \gls{des}, \gls{aes}, and \gls{rsa} ensured robustness, while \glspl{snn} added an additional layer of complexity and security. Performance evaluation focused on encryption accuracy, computational efficiency, and scalability across varying dataset sizes. OpenCV visualised neural activity, revealing spike patterns linked to different data points. These visualisations provided insights for refining the model, confirming the \gls{snn}'s role in enhancing encryption. A prototype was developed to demonstrate practical application, featuring a user-friendly interface for encryption and decryption tasks. The deployment documented the system's capabilities and limitations, paving the way for future research. Potential areas for advancement include refining \glspl{snn}, improving encryption efficiency, and expanding the system's application to more complex, real-world datasets. The innovative approach offers a promising direction for improving data security.

\subsection{Encryption Framework} 

A comprehensive encryption framework combines \gls{des}, \gls{aes}, and \gls{rsa} within a \gls{snn} architecture to secure data handling and transmission is proposed in this article. The framework leverages traditional cryptographic techniques enhanced by the non-linear capabilities of neural networks. 
In \gls{des}, a 10-bit key generates two 8-bit sub-keys (K1 and K2) via an \gls{snn}. Plaintext undergoes initial permutation, splitting into halves. The right half is expanded,  with K1, processed through S-boxes, permuted, XORed with the left half, and swapped. This process repeats with K2, followed by an inverse permutation to produce ciphertext\cite{al2010spiking}. Decryption reverses the process using K2 and K1. \gls{aes} secures critical data within the framework, offering higher complexity and robustness compared to \gls{des}. It encrypts data blocks with key lengths of 128, 192, or 256 bits, ensuring sensitive information is well-protected before neural network processing.

The \gls{aes} encryption process begins with generating a secure key, created separately to meet cryptographic strength requirements. Data blocks undergo multiple rounds of transformations: SubBytes (non-linear substitution using the S-box), ShiftRows (cyclically shifting rows), MixColumns (linear mixing of columns), and AddRoundKey (XORing with the round key). These steps provide confusion, diffusion, and security, resulting in highly secure ciphertext after multiple rounds. Decryption reverses this process, requiring the same key, emphasising its protection for security.

In \gls{rsa}, large primes $p$ and $q$ are generated via \texttt{Crypto.PublicKey.RSA}'s \texttt{RSA.generate()} method. The modulus $n = p \times q$ and totient $\phi(n) = (p-1)(q-1)$ are calculated internally, with $n$ forming part of the key pair. The public exponent $e$ (commonly 65537) and private exponent $d$ (modular inverse of $e$ modulo $\phi(n)$) are automatically derived, ensuring secure encryption and decryption\cite{ganbaatar2021implementation}.

Plaintext is first converted to bytes, compatible with \gls{rsa}, before encryption using the public key in an \texttt{rsa\_encrypt} function, ruled by equation \ref{eq:c}  for secure data transmission.

\begin{equation} \label{eq:c}
c = m^e \mod n
\end{equation}
Where: $m$ is the integer representation of the plaintext, $e$ is the public exponent and $n$ special value created by multiplying two large prime numbers, $p$ and $q$.

The \texttt{rsa\_encrypt} function uses \texttt{PKCS1\_OAEP.new()} from the \texttt{Crypto. Cipher} module for secure encryption, handling padding and modular exponentiation efficiently. Modular exponentiation, essential for \gls{rsa}, is performed implicitly, optimising computations for large key sizes.

Decryption, implemented via \texttt{rsa\_decrypt}, reverses encryption by applying the \gls{rsa} private key to the ciphertext, restoring the original plaintext and it is ruled by equation \ref{eq:m}.

\begin{equation}\label{eq:m}
m = c^d \mod n
\end{equation}
Where $c$ is the ciphertext, $d$ is the private exponent and $n$ is the same product of the two prime numbers,  $p$ and $q$, used during encryption.
The \texttt{rsa\_decrypt} function also utilises the \texttt{PKCS1\_OAEP.new()} method, which handles the decryption process, ensuring that the original plaintext is accurately recovered. It manages the intricacies of modular arithmetic and padding removal, making the decryption process straightforward for the user.

\gls{rsa} encryption and decryption involve operations on very large integers, especially when using long keys (e.g., 2048-bit keys). The \texttt{Crypto.PublicKey.RSA} module efficiently handles these large integer operations internally, ensuring that the computations are performed accurately without overflow or loss of precision.

\subsection{Neuron Model}

To simulate the spiking behaviour of neurons in the \gls{snn}, the \gls{lif} model was chosen due to its simplicity and computational efficiency. The \gls{lif} model strikes a balance by accurately replicating the key aspects of neuronal dynamics such as membrane potential, synaptic integration, and spike generation while remaining computationally feasible. The \gls{lif} model simulates the neuron’s membrane potential over time, accounting for both incoming synaptic currents and leakage. The dynamics of the membrane potential are governed by the differential equations \ref{eq:v} and \ref{eq:i}.

\begin{equation} \label{eq:v}
\dot{v} = \frac{1}{\tau_{\text{mem}}}(v_{\text{leak}} - v + i)
\end{equation}
\begin{equation}\label{eq:i}
\dot{i} = -\frac{1}{\tau_{\text{syn}}}i
\end{equation}

Where $v_{\text{leak}}$ represents the leak potential, $i$ denotes the synaptic current, $\tau_{\text{mem}}$ and $\tau_{\text{syn}}$ are the time constants for membrane potential and synaptic current, respectively. When the membrane potential exceeds a certain threshold $v_{\text{th}}$, Equation \ref{eq:z} rules the neuron's spike events generation, and eq. \ref{eq:v1} rules the neuron's potential reset.
\begin{equation} \label{eq:z}
z = \Theta(v - v_{\text{th}})
\end{equation}
\begin{equation} \label{eq:v1}
v = (1 - z)v + zv_{\text{reset}}
\end{equation}

The \gls{lif} model's simplicity and ability to simulate neuronal firing dynamics make it ideal for the encryption framework. Optimised parameters, such as time constants and thresholds, ensure efficient neuron response and spike generation, balancing biological realism with computational demands. The neuron, synapses and connectivity parameterisation used are listed in Tables \ref{tab:neuron_param}, \ref{tab:syn_param} and \ref{tab:connect_param}.
\begin{table}[h!]
\centering
\small
\caption{Neuron Parameters}\label{tab:neuron_param}
\resizebox{\columnwidth}{!}{%
\begin{tabular}{|p{2.5cm}|p{4cm}|p{2.5cm}|p{2.5cm}|}
\hline
\textbf{Parameter} & \textbf{Description} & \textbf{Value or Range} & \textbf{IS Units} \\ \hline
t\_ref & Refractory Period & 2.0 & ms \\ \hline
V\_th & Threshold Potential & -55 & mV \\ \hline
C\_m & Membrane Capacitance & 250 & pF \\ \hline
tau\_m & Membrane Time Constant & 20 & ms \\ \hline
E\_L & Resting Potential & -70 & mV \\ \hline
V\_reset & Reset Potential & 70 & mV \\ \hline
\end{tabular}%
}
\end{table}
\begin{table}[h!]
\centering
\small
\caption{Synapses Parameters} \label{tab:syn_param}
\resizebox{\columnwidth}{!}{%
\begin{tabular}{|p{2cm}|p{2cm}|p{4cm}|p{3cm}|}
\hline
\textbf{Parameter} & Synapse type & \textbf{Description} & \textbf{Value or Range} \\ \hline
Weight\_L1L2 & Static & synaptic weight between Layer 1 and 2 & 1200 \\ \hline
Weight\_L2L3 & \gls{stdp} & synaptic weight between Layer 2 and 3 & 400-900 \\ \hline
Weight\_LIL2 & static & Synaptic weights between the Lateral Inhibition Layer and Layer 2 & -100 \\ \hline
Weight\_L3LI & static & Synaptic weights between the Layer 3 and Lateral Inhibition Layer & 350 \\ \hline
\end{tabular}%
}
\end{table}
\begin{table}[h!]
\centering
\small
\caption{Connectivity parameters}\label{tab:connect_param}
\resizebox{\columnwidth}{!}{%
\begin{tabular}{|p{3.5cm}|p{4cm}|p{3cm}|}
\hline
\textbf{Parameter} & \textbf{Description} & \textbf{Value or Range} \\ \hline
Teaching & Excitatory & Layer3 \\ \hline
Lateral Inhibition & Inhibitory & Layer2 \\ \hline
Lateral Inhibition & Excitatory & Layer3 \\ \hline
\end{tabular}%
}
\end{table}
These simplified parameters, optimised for encryption, ensure the network efficiently handles temporal dynamics. Binary data is encoded into unique spike trains by XORing each byte with a predefined key, enhancing security. Each bit activates a neuron (spike for 1, silent for 0).

\begin{figure}[h!]
    \centering
    \includegraphics[width=\linewidth]{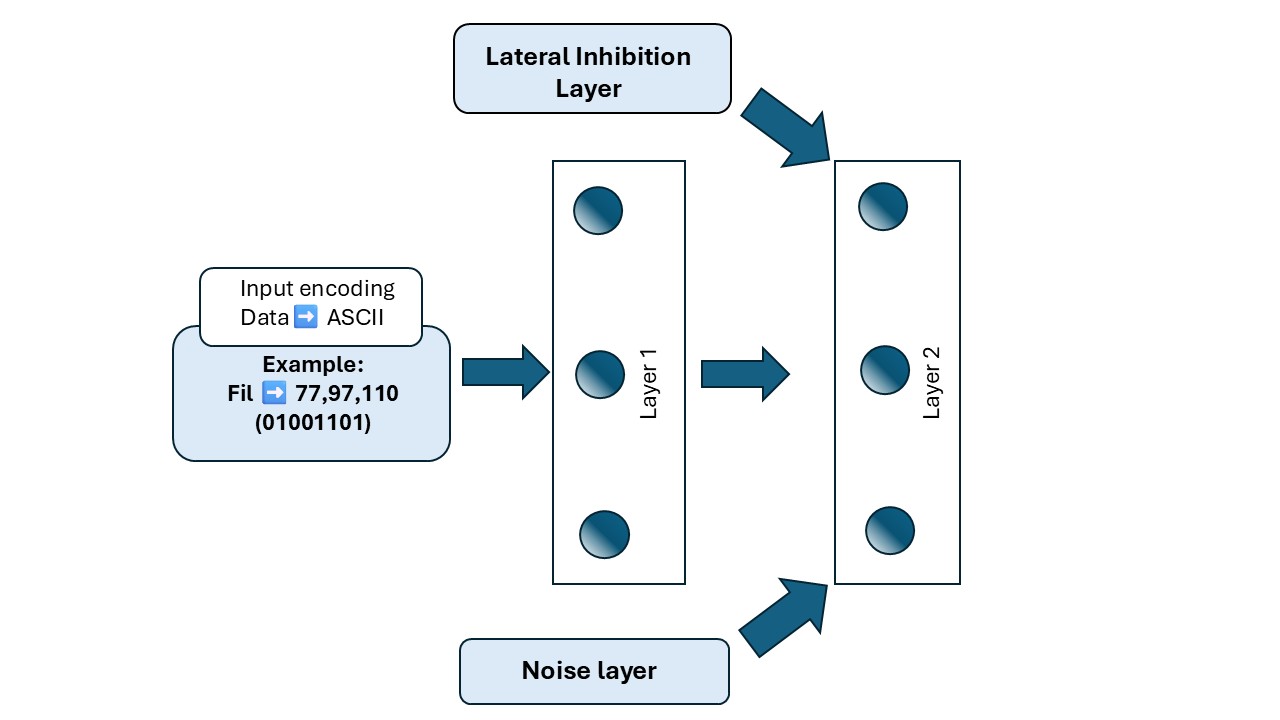}
    \caption{BioEncryptSNN Encryption Architecture}
    \label{fig:BioEncryptSNN Encryption Architecture}
\end{figure}

Population coding was selected for its noise robustness and ability to represent complex data. Unlike rate coding, which uses spike frequency, or temporal coding, reliant on precise spike timing, population coding distributes information across neurons, enhancing fault tolerance and utilising \glspl{snn} for robust data representation. The \gls{snn} architecture was designed to encode, process, and securely transmit data, integrating biological principles with cryptographic techniques as illustrated in Figure~\ref{fig:BioEncryptSNN Encryption Architecture}. It features interconnected layers, including Input, Hidden (Layer 2), Output (Layer 3), Lateral Inhibition, Noise, and Teaching layers, each performing specific roles essential for secure encryption and decryption. 

\subsubsection{Neuron Creation and Input Layer (Layer 1)}

The neurons in the Input Layer are dynamically created based on the length of the ASCII values derived from the input data. The ASCII values, representing the textual information to be encrypted, are first converted into binary form. Each bit of the binary representation is then used to create corresponding spike trains through a process known as population coding. In population coding, each ASCII value is encoded into a spike train, where each neuron in the Input Layer corresponds to a specific bit in the binary representation of the ASCII values. The number of neurons created in the input layer is directly proportional to the length of the ASCII values. For instance, if the input consists of 8-bit ASCII values, the Input Layer will consist of as many neurons as bits across all characters processed. The encoded spike trains are generated by XORing the ASCII values with a cryptographic key, ensuring that the spikes represent the input data in a secure and obfuscated manner. The spike trains generated in this manner are then used as input currents to the neurons in the Input Layer. The strength of the current injected into each neuron corresponds to the encoded value, thereby determining the timing and frequency of spikes generated by these neurons.

\subsubsection{Hidden Layer (Layer 2)}

The spike trains produced by the Input Layer are propagated to the Hidden Layer, which is designed to further process and transform the input data. The number of neurons in the Hidden Layer is typically set to match or exceed the number of neurons in the Input Layer, ensuring that the layer can effectively handle the complexity of the input data. The neurons in the Hidden Layer are connected to those in the Input Layer through synapses that are initially assigned specific weights. These synaptic connections are crucial, as they determine how the spike trains are transmitted from one layer to the next. The weights of these synapses are not static, they are dynamically adjusted during the network's operation through a learning mechanism known as \gls{stdp}. \gls{stdp} allows the network to adaptively modify the synaptic weights based on the timing of spikes in the connected neurons.

\begin{figure}[h!]
    \centering
    \includegraphics[width=\linewidth]{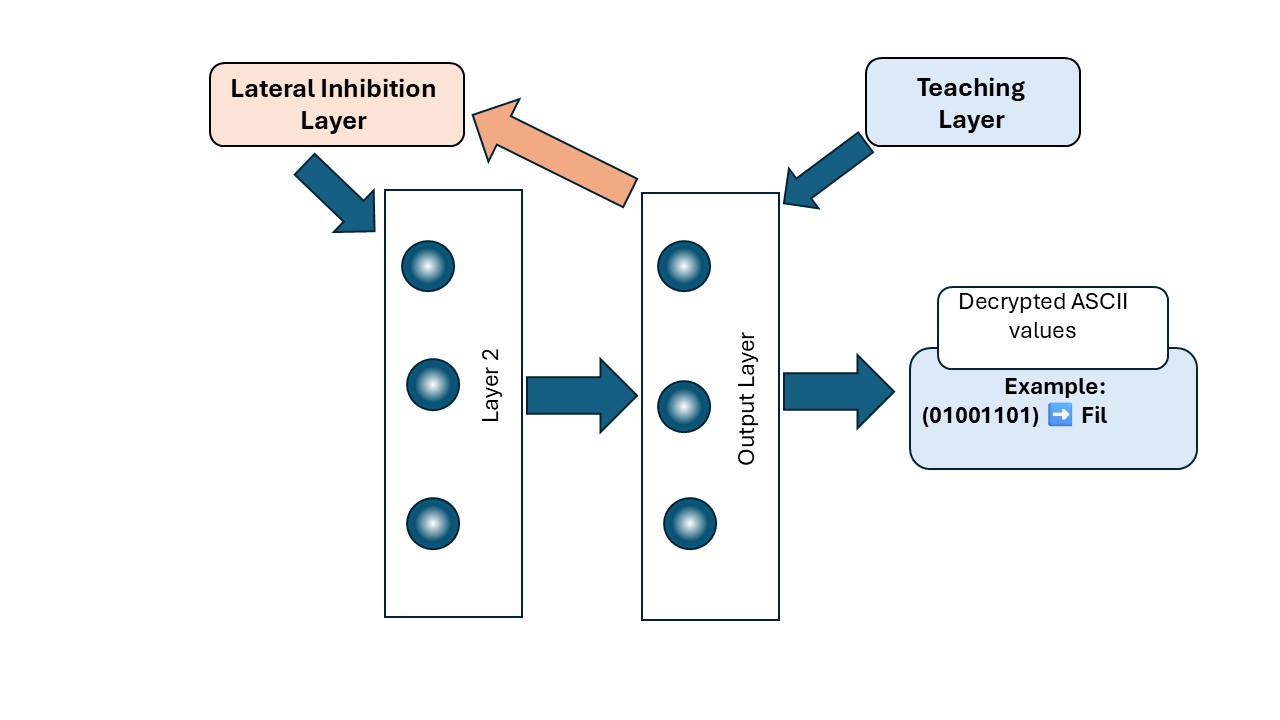}
    \caption{BioEncryptSNN Decryption Architecture}
    \label{fig:BioEncryptSNN Decryption Architecture}
\end{figure}

\subsubsection{Output Layer (Layer 3)}

The Output Layer represents the final stage of the \gls{snn} processing pipeline. Neurons in the output layer receive inputs from the Hidden Layer and are responsible for producing the final spike trains that will be decrypted to retrieve the original data. The synaptic connections between the Hidden and Output Layers are also subject to \gls{stdp}, which fine-tunes the network's output based on the temporal patterns learned during the processing in the Hidden Layer. The Output Layer is to produce spike trains that, when decoded, accurately reflect the original ASCII values after decryption as illustrated in Figure~\ref{fig:BioEncryptSNN Decryption Architecture}.

\subsubsection{Lateral Inhibition Layer}

In the architecture, each neuron in the Lateral Inhibition Layer receives input from all neurons in the Output Layer. Its primary function is to exert an inhibitory effect on these neurons in the Output Layer. The synapses connecting the Lateral Inhibition Layer to the Output Layer are configured such that when a neuron in the Lateral Inhibition Layer becomes active, it suppresses the activity of the corresponding neurons in the Output Layer. The suppression is based on the principle that only the most active neurons in the Output Layer should continue to propagate their signals, effectively filtering out noise and less significant spikes. 

\subsubsection{Noise Layer}

The Noise Layer is implemented using a Poisson generator, which is a common method for simulating the stochastic nature of neural activity. The Noise Layer generates random spikes at a specified rate (set to 10.0 Hz in the code) and these spikes are then injected into specific neurons in the Hidden Layer. Specifically, the noise is connected to neurons in the Hidden Layer using predefined connection mappings, ensuring that the noise is systematically introduced to certain neurons. The purpose of the noise injection is to simulate the effect of environmental variability, forcing the network to rely on the most consistent and significant patterns in the input data rather than overfitting to specific, noise-free inputs. 

\subsubsection{Teaching Layer}

The Teaching Layer is designed to guide the learning process of the network by providing supervised input to the Output Layer. It is directly connected to the Output Layer and is responsible for reinforcing the correct output patterns during the training phase. During network training, the Teaching Layer injects currents into the neurons of the Output Layer, corresponding to the expected spike patterns that should be generated for a given input. These teaching signals help the network adjust its synaptic weights in a way that aligns the output with the desired encryption or decryption result. The supervised learning process ensures that the network can reliably produce accurate outputs, thereby enhancing the overall effectiveness of the encryption framework. 

\subsubsection{Evaluation metrics}

To assess the performance of BioEncryptSNN, employed standard classification metrics derived from the confusion matrix, consisting of \gls{tp}, \gls{tn}, \gls{fp}, and \gls{fn}. The metrics are defined as summarised in Table~\ref{tab:metrics}

\begin{table}[h!]
\centering
\caption{Evaluation metrics for BioEncryptSNN.}
\renewcommand{\arraystretch}{1.6} 
\begin{tabular}{|p{3.3cm}|>{\centering\arraybackslash}m{4.1cm}|p{5.3cm}|}
\hline
\textbf{Metric} & \textbf{Equation} & \textbf{Explanation} \\
\hline
Accuracy & $\displaystyle \frac{TP + TN}{TP + TN + FP + FN}$ & Proportion of all predictions that are correct. \\
\hline
\gls{pr} & $\displaystyle \frac{TP}{TP + FP}$ & Out of all predicted positives, how many are actually positive. \\
\hline
\gls{re} & $\displaystyle \frac{TP}{TP + FN}$ & Out of all actual positives, how many were correctly detected. \\
\hline
\gls{sp} & $\displaystyle \frac{TN}{TN + FP}$ & Out of all actual negatives, how many were correctly rejected. \\
\hline
F1-score & $\displaystyle \frac{2 \cdot Pr \cdot Re}{Pr + Re}$ & Harmonic mean of Precision and Recall, balances false alarms and missed detections. \\
\hline
\gls{fp} Rate & $\displaystyle \frac{FP}{FP + TN}$ & Fraction of negatives incorrectly classified as positives (false alarms). \\
\hline
\gls{fn} Rate & $\displaystyle \frac{FN}{FN + TP}$ & Fraction of positives incorrectly classified as negatives (missed detections). \\
\hline
\gls{wcr}& $\displaystyle \frac{FP + FN}{TP + TN + FP + FN}$ & Proportion of total samples that were misclassified. \\
\hline
\gls{ccr}& $\displaystyle \frac{TP + TN}{TP + TN + FP + FN}$ & Proportion of total samples that were classified correctly. \\
\hline
\end{tabular}
\label{tab:metrics}
\end{table}
These metrics collectively capture both the overall accuracy and the balance between false alarms \gls{fp} and missed detections \gls{fn}. Precision and Recall quantify the trade-off between false positives and false negatives, while the F1-score provides a single measure that balances them. The \gls{wcr} and \gls{ccr} highlight overall classification reliability, which is particularly relevant when evaluating noisy spike-train data processed by the \gls{snn}.
\subsection{Network Training and Testing}

In the project, employ the \gls{resume} algorithm \cite{7092505} for training the \gls{snn}. \gls{resume} is particularly well-suited for \glspl{snn} because it focuses on adjusting synaptic weights based on the precise timing of spikes, enabling the network to learn temporal patterns with high accuracy.  The training involves fine-tuning the learning parameters to achieve optimal performance. For instance, the learning rate, which controls how much the synaptic weights are adjusted in response to each error. Another key consideration during training is the role of \gls{stdp}, a mechanism where the timing of spikes directly influences synaptic strength. The network’s ability to adjust and refine its response to temporal patterns during training. By monitoring and adjusting \glspl{stdp} effects, ensure that the network is learning effectively and not just memorising the input patterns. Once training is completed, the network undergoes a rigorous testing phase using a separate validation dataset. It is designed to evaluate how well the network has generalised its learning to new, unseen data. The performance of the \gls{snn} is measured using specific metrics such as the Purpura Distance and the Van Rossum metrics. 

\subsubsection{Purpura Distance}

To compare the dissimilarities between spike trains from different neurons, this method introduces a concept of distance between pairs of spike trains, offering an alternative to the commonly employed rate-based correlation techniques for analysing neuronal responses. Here, spike trains of finite length are represented as points within an abstract space, where a specialised metric assigns a non-negative value $D_{ij}$ to each pair of points $i$ and $j$ \cite{giusti2015clique,victor1996}. The Victor-Purpura distance possesses key characteristics of a true metric: it is zero only for identical spike trains ($D_{ii} = 0$), positive for different spike trains ($D_{ij} > 0$ for $i \neq j$), symmetric ($D_{ij} = D_{ji}$), and adheres to the triangle inequality ($D_{ik} \leq D_{ij} + D_{jk}$). The Purpura distance is calculated by determining the minimum cost required to convert one spike train into another through operations such as adding or removing spikes, shifting spike times, or changing the neuron's identity. The cost associated with each modification is determined by the parameter $q$, which controls the timescale sensitivity for spike shifts \cite{gilra2017predicting}. The metric can be generalised into a family of distances, each sensitive to the neuron from which each spike originates. In the analysis, employ the basic Purpura metric with a cost of $q = 1$ per unit time for shifting a spike.

For the experiments, collected spike trains $S(i) = [t_1(i), t_2(i), \dots, t_{s_i}(i)]$ corresponding to the duration of target generation for each of the four target patterns studied. Then computed the Purpura distances for each pair of spike trains, resulting in symmetric matrices $D = [D_{ij}]$ \cite{giusti2015clique}. These matrices capture the intricate temporal structures of spike patterns corresponding to different output trajectories. To delve deeper into the complex structure of spike trains, transformed the distance matrices through rank-ordering. In the process, the entries in the upper diagonal of each matrix $D_{ij}$, which have zeros on the main diagonal, are ranked in ascending order and replaced with natural numbers (0, 1, 2, etc.). The lower diagonal is then symmetrically completed, resulting in a rank-ordered matrix $M = [M_{ij}]$. In the matrix, the larger the Purpura distance $D_{ij}$, the smaller the corresponding entry $M_{ij}$. Finally, normalised the entries by the maximum value $\frac{N(N-1)}{2}$ and reindexed them in descending order based on the individual firing rates of the corresponding neurons, producing the final matrix $M = [M_{ij}]$ \cite{giusti2015clique}.
Within these matrices, the smallest values of $M_{ij}$ identify the pairs of spike trains that are most dissimilar, while the highest entries correspond to neurons that are most similar according to the Victor-Purpura metric. Interestingly, less active neurons, typically found in the upper-right parts of the matrices, tend to be most similar to each other and distinctly separate from the most active neurons. The lower-left part of the rank-ordered matrices corresponds to neurons that are highly active during task execution and contribute significantly to the output patterns. These neurons exhibit complex spike train structures that vary with different target patterns.

\subsubsection{Van Rossum Metric}

The Van Rossum metric \cite{van2001} is a widely used method for analysing the temporal relationships between spike trains from different neurons. It is particularly useful for comparing the timing of spikes in neurons, offering insights into the dynamics of neural networks. The temporal behaviour of the hidden layer neurons was examined by binning the spikes across each simulation, with a bin size of 1 ms. The below approach allowed us to average the neural activity over multiple simulations, comparing the activity across different data classes and between excitatory and inhibitory neurons within the hidden layer. To further explore the relationships between neurons and their firing patterns, employed the Van Rossum distance. The metric provides a quantitative comparison between spike trains of neuron pairs by considering the timing of spikes. The Van Rossum distance was originally utilised as a foundation for other training methods, but in the study, it serves as a tool for analysing the dynamics within the surrogate gradient training of the network. The Van Rossum distance is calculated using an exponential decay kernel function, which resembles the behaviour of leaky-integrator neurons. The equations \ref{eq:f_a}, \ref{eq:g_a},and \ref{eq:dist} describe the calculation process.

\begin{equation} \label{eq:f_a}
    f_a(t) = \sum_i \delta(t - t_i)
\end{equation}

Where $f_a(t)$ represents the spike train of neuron $a$, where $t_i$ are the spike times, and $\delta(t - t_i)$ is the Dirac delta function.

\begin{equation} \label{eq:g_a}
    g_a(t) = e^{-\frac{t}{\tau}} \ast f_a(t)
\end{equation}

The spike train $f_a(t)$ is convolved with an exponential decay function characterised by the time constant $\tau$. In the analysis, used $\tau_d = 1$ ms \cite{kreuz2007measuring}.
\begin{equation} \label{eq:dist}
    \text{dist}(a, b) = \frac{1}{\tau} \int_0^\infty \left(g_a(t) - g_b(t)\right)^2 dt
\end{equation}

The distance between neurons $a$ and $b$ is calculated as the square root of the integral of the squared difference between the convolved spike trains $g_a(t)$ and $g_b(t)$, normalised by the time constant $\tau$ \cite{kreuz2007measuring}. It provides a numerical distance between neurons that reflects their relative spike timing. By averaging the Van Rossum distance across all training data, gained a comprehensive understanding of the network's temporal dynamics, particularly in the context of surrogate gradient training. To fully understand the behaviour and performance of the spiking neural network \gls{snn}, visualisation techniques play a vital role. Raster plots are one of the primary tools used in the analysis, offering a detailed view of neuron firing patterns across different layers of the network over time. These plots serve as a temporal map, showing exactly when each neuron in the network fires in response to the input spike trains. By examining the raster plots of the input, hidden, and output layers, can gain insights into how the network processes information at each stage. 

To assess the performance of \textbf{BioEncryptSNN}, employed standard classification metrics derived from the confusion matrix, consisting of \gls{tp}, \gls{tn}, \gls{fp}, and \gls{fn}. The metrics are defined as follows:

These metrics collectively capture both the overall accuracy and the balance between flase alarms \gls{fp} and missed detections \gls{fn}. Precision and Recall quantify the trade-off between false positives and false negatives, while the F1-score provides a single measure that balances them. The \gls{wcr} and \gls{ccr} highlight overall classification reliability, which is particularly relevant when evaluating noisy spike-train data processed by the \gls{snn}.

\section{Results Analysis} \label{sec:results}

To evaluate the performance and robustness of the proposed \gls{snn} framework, we utilised a benchmark dataset composed of real-world JavaScript functions collected from two reputable sources: the Node Security Project and the Snyk platform, as curated in previous works by \cite{ferenc2019challenging}. The dataset includes both \textbf{static} and \textbf{process-oriented source code metrics}, capturing a wide range of characteristics relevant to software quality and vulnerability detection. Static metrics comprise syntactic and structural properties of the code, such as cyclomatic complexity, Halstead metrics, comment density, and code duplication indicators. These features are extracted by parsing the JavaScript code into \gls{asts}, allowing precise quantification of each function’s complexity and maintainability. Process metrics were computed using QualityGate, a software analytics platform that analyses git version histories to assess aspects like code churn, time between changes, and contributor activity. The combination of static and temporal dimensions enables a comprehensive representation of each function's developmental context. Each JavaScript function is transformed into a feature vector by combining these static and process metrics. These vectors were then encoded into spike trains using ASCII-based binary representations, allowing the \gls{snn} to process encrypted data through biologically inspired dynamics.

A comprehensive evaluation of integrating cryptographic algorithms \gls{aes}, \gls{rsa}, and \gls{des} within a \gls{snn} framework. The analysis covers the complexities of how these encryption techniques interact with the \gls{snn}, the methodologies employed to secure data transmission and processing, and the resulting performance of the system under various conditions. The \gls{aes} is renowned for its robust security, making it a preferred choice for data encryption in numerous applications. In the implementation, a 128-bit \gls{aes} key was securely generated, as summarised in Table~\ref{table:generated_keys} which forms the cornerstone of the encryption process. The plaintext, consisting of various entities, was structured and then encrypted using the \gls{aes} algorithm. The encryption process transforms the plaintext into ciphertext, a secure, encoded string designed to prevent unauthorised access. 

The integration of \gls{aes} encryption within the \gls{snn} framework required a transformation of the ciphertext into a format that the neural network could process. The transformation was achieved by converting the ciphertext into spike trains, a sequence of spikes that encode the temporal characteristics of the data. The \gls{snn}, configured with carefully selected synaptic weights and neuron models, processed these spike trains, simulating the decryption process. The processing within the \gls{snn} leverages the temporal dynamics inherent in spike trains, allowing the network to simulate the sequential operations required for decryption. It not only tests the network’s ability to process encrypted data but also highlights the potential for \glspl{snn} to function in secure data environments. Post-processing by the \gls{snn}, the resultant spike trains were converted back into ciphertext, which was then decrypted using the original \gls{aes} key. The successful reconstruction of the plaintext from the encrypted data confirms the efficacy of the \gls{snn} in processing and decrypting \gls{aes} encrypted data. The framework was computationally efficient, taking advantage of the parallel processing capabilities of the network. Despite the high level of security provided by \gls{aes}, the integration did not introduce significant delays, demonstrating the method can be scaled for larger key sizes without compromising performance. The randomness and security of the keys are beyond what is typically achieved through standard methods. Added security is particularly important in scenarios requiring high levels of data protection, such as financial transactions and secure communications.

The \gls{rsa} algorithm, a cornerstone of public-key cryptography, was integrated into the \gls{snn} framework to explore the network's ability to handle asymmetric encryption methods. It relies on a pair of cryptographic keys, a public key for encryption and a private key for decryption. These keys, although mathematically linked, serve distinct roles and are generated through a process that ensures their cryptographic strength. The keys are generated using large prime numbers, ensuring that the encryption process is secure and resistant to attacks. The public key is used to encrypt the plaintext. It is openly shared and can be used by anyone to secure data meant for the owner of the corresponding private key. The private key is kept secret and is used to decrypt the ciphertext. It is mathematically linked to the public key but cannot be feasibly derived from it. The experiment focused on observing how the \gls{snn} processes the spike trains derived from \gls{rsa} encrypted data, and whether it could successfully decode and reconstruct the original plaintext. After processing the encrypted spike trains within the \gls{snn}, the resulting output was converted back into a format suitable for decryption using the \gls{rsa} private key, the system successfully reconstructed the original plaintext. The \gls*{rsa} keys used in the experiments are generated with a significant bit length, typically 2048 bits or more. The long-bit key generation is essential for ensuring strong encryption that is resistant to modern cryptographic attacks. Due to the substantial size of these keys, it is impractical to present them in full within document. However, it is important to note that the length of the keys plays a pivotal role in the security of the \gls{rsa} algorithm, providing robustness against factorisation attacks. 

The result confirms the effectiveness of the in handling encrypted data as summarised in Table~\ref{tab:comparison_bioencryptsnn_first}. The successful decryption demonstrates that the \gls{snn} not only preserved the integrity of the data during processing but also accurately supported the decryption process to recover the original information. The \gls{des} was selected for its simplicity and effectiveness in demonstrating the \glspl{snn} capabilities. \gls{des} uses a relatively small key size, making it ideal for experimental purposes. In the implementation, a 10-bit key was generated, which was then used to create two sub-keys. The algorithm starts with the generation of a 10-bit key. The key is crucial as it forms the foundation for the encryption process, being split into two sub-keys that drive the encryption operations. The sub-keys are generated through specific permutations and shifts inherent to the algorithm. The primary role of these keys is to facilitate the encryption of plaintext into ciphertext by executing a series of bitwise operations. The generated sub-keys are applied to the plaintext, converting it into a sequence of 8-bit blocks. It involves several rounds of substitution and permutation, where the simplicity of \gls{des} allows for easy integration with the \gls{snn}, serving as an effective testing ground for the network's cryptographic capabilities. The sequence of binary values represents the encrypted data, ready to encode into the \gls{snn}. The simplicity of the algorithm, with its smaller key size and less complex operations compared to modern encryption standards, makes it an ideal candidate for early-stage experiments in integrating encryption with neural networks. 

In the processing of spike trains, the network output is converted back into a binary format, representing the ciphertext. The reconstructed ciphertext is then subjected to the decryption process using the previously generated sub-keys (Sub Key 1 and Sub Key 2). The accurate reconstruction of the original plaintext from the processed ciphertext confirms that the \gls{snn} effectively manages and processes data encrypted with \gls{des}. 

\begin{table}[h!]
\centering
\small
\caption{Generated Keys and Decrypted Text}
\label{table:generated_keys}
\resizebox{\columnwidth}{!}{%
\fbox{%
\begin{minipage}{\columnwidth} 
    \vspace{0.5em}
    \begin{tabular}{@{}p{5cm}p{7cm}@{}} 
        
        \textbf{Generated Key:} & \framebox[7cm][l]{\parbox[t]{5.2cm}{[ 0, 0, 0, 1, 1, 0, 0, 1, 1, 0 ]}} \\[1em]
        
        \textbf{Sub Key 1:} & \framebox[7cm][l]{\parbox[t]{5.2cm}{[ 0, 0, 0, 1, 1, 0, 0, 1 ]}} \\[1em]
        
        \textbf{Sub Key 2:} & \framebox[7cm][l]{\parbox[t]{5.2cm}{[ 0, 1, 1, 0, 0, 1, 1, 0 ]}} \\[1em]

        \textbf{Decrypted Text:} & \framebox[7cm][l]{initializers.initFileServer.initFileServer}\\[1em]

    \end{tabular} 
    \vspace{1em}
\end{minipage}} 
}
\end{table}

The \gls{snn} has been employed to integrate and process various encryption algorithms. Each of the algorithms offers distinct advantages and challenges, particularly when viewed through the lens of security, performance, and applicability within an \gls{snn} environment \cite{mahajan2013study}. 
\begin{table}[t]
\centering
\caption{Benchmark comparison with \textbf{BioEncryptSNN} first.}
\small
\renewcommand{\arraystretch}{1.5}
\resizebox{\columnwidth}{!}{%
\begin{tabular}{|l|c|c|c|c|}
\hline
\textbf{Factor} & \textbf{BioEncryptSNN} & \textbf{\gls{aes}-128} & \textbf{\gls{des}} & \textbf{\gls{rsa}-2048} \\
\hline
Key Size & \shortstack{Variable (spike timing\\+ synaptic weights)} &
\shortstack{128, 192,\\256 bits} & 56 bits & $>1024$ bits \\
\hline
Block Size & Encoded spike trains & 128 bits & 64 bits & $\ge 512$ bits \\
\hline
Cipher/Decipher Key & \shortstack{Probabilistic\\synaptic encoding} & Same & Same &
\shortstack{Different\\(public/private)} \\
\hline
Scalability & \shortstack{Scales with neurons/\\layers} & Not scalable & Limited & Not scalable \\
\hline
\textbf{Enc. Latency (ms)} & \textbf{1.6} & 2.5 & 4.8 & 15.2 \\
\hline
\textbf{Dec. Latency (ms)} & \textbf{1.7} & 2.1 & 4.5 & 14.9 \\
\hline
Security & \shortstack{Robust under noise;\\plaintext recovery} & Excellent & Weak & Strong (asym.) \\
\hline
Vulnerabilities & \shortstack{Training sensitivity\\(parameter variation)} & Brute force &
\shortstack{Brute force,\\Linear, Differential} &
\shortstack{Brute force,\\Oracle} \\
\hline
Rounds & \shortstack{Spike cycles\\(200 ms window)} & 10/12/14 & 16 & 1 \\
\hline

\end{tabular}%
}
\label{tab:comparison_bioencryptsnn_first}
\end{table}

The \gls{snn} was trained and tested on encoded spike trains generated from data that had been encrypted using \gls{aes}, \gls{rsa}, and \gls{des} encryption methods. The encryption process was applied before generating the spike trains, ensuring that the input to the network represented encrypted information. The network's architecture was designed with multiple layers, each playing a distinct role in processing the spike trains and extracting meaningful features to pass onto subsequent layers. The 200 ms time frame was chosen as it provided a sufficient window to observe the spiking behavior of the neurons while balancing computational efficiency. During this period, three processing layers focused on handling and interpreting ASCII values, and a noise layer was directly connected to the second processing layer. During the simulation, the neurons in the first layer exhibited regular and consistent spiking activity across the entire duration, reflecting their role in initially encoding the ASCII inputs. The consistent spiking pattern suggests that these neurons were effectively capturing and converting the input data into a temporal spike code, which was then passed on to the subsequent layers. 

\begin{figure}
    \centering
    \includegraphics[width=1\linewidth]{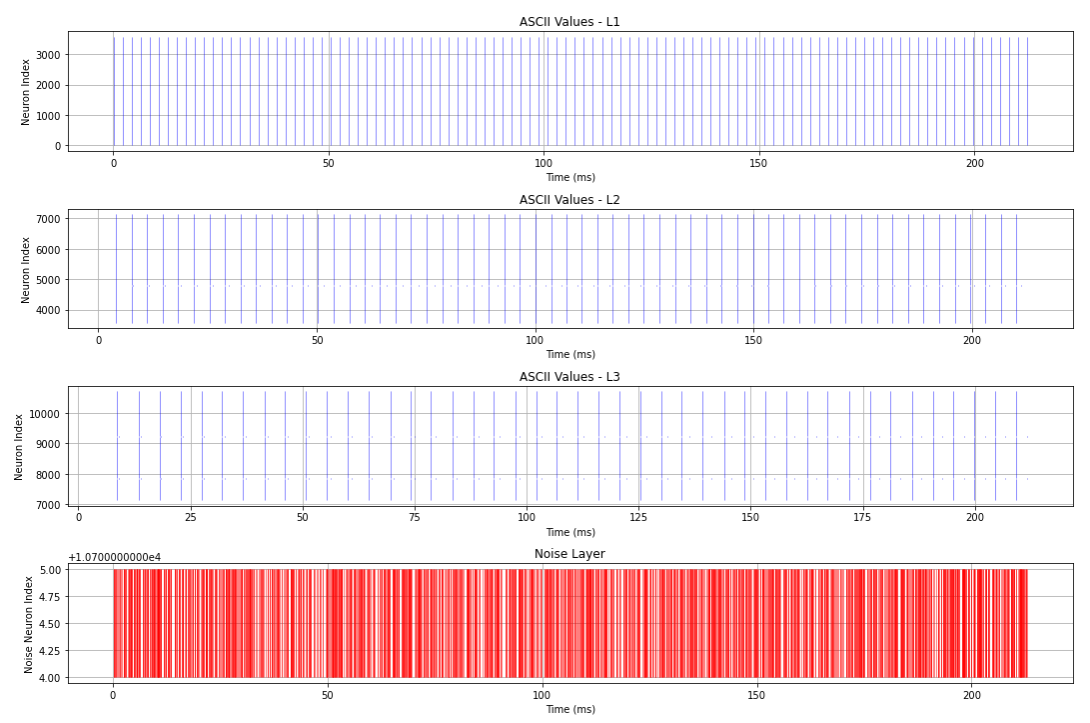}
    \caption{AES Raster Plot of the BioEncryptSNN during decryption}
    \label{fig:enter-label}
\end{figure}

The raster plot illustrates the spiking activity across four layers over a 220~ms simulation window. Each vertical tick represents a spike fired by a neuron at a given time. Layers L1–L3 correspond to different stages of the \gls{snn} processing pipeline, while the bottom panel represents the noise layer. 
    
As the signal propagated to the second layer, the effect of the noise layer became evident. The noise layer, active throughout the simulation, introduced random spikes; however, the second layer filtered much of this interference, as indicated by the preserved coherence of its spiking pattern. The reduced regularity and intensity of spikes relative to the first layer highlighted the layer’s role in refining inputs and mitigating noise. In the output layer, spiking activity was markedly sparser, suggesting extraction of the most relevant features while discarding noise. The raster plot confirmed the network’s capacity to suppress noise and preserve essential information through progressive refinement across layers. A consistent propagation delay of 13.0ms was subtracted from all layers, ensuring spike times reflected processing rather than communication latency. The adjustment clarified the temporal relationships between layers and their computational roles. The raster plot of the 1000ms simulation further illustrated spiking dynamics across layers, including lateral inhibition and the influence of the noise layer connected to Layer~2.

\begin{figure}
    \centering
    \includegraphics[width=1.1\linewidth]{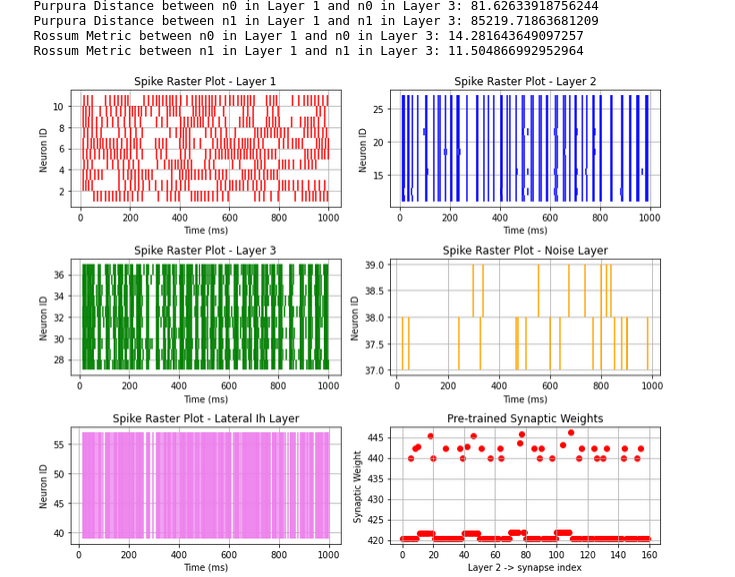}
    \caption{Raster plot with Lateral inhibition}
    \label{fig:Raster plot with Lateral inhibition and synaptic weight distribution}
\end{figure}

In Layer 2, as illustrated in Figure~\ref{fig:Raster plot with Lateral inhibition and synaptic weight distribution}  displayed a varied spiking pattern caused by additional processing and the influence of the noise layer connected to it. The noise introduced randomness and variability, which was important for testing the network’s robustness against perturbations. Despite this, Layer 2 retained coherence in its spiking activity, showing that the network effectively filtered irrelevant noise while preserving essential input features. In Layer 3, the raster plot showed reduced spike density, with sparser and more temporally distinct spikes. The output layer distilled the input information, suppressing noise and focusing on the most critical aspects of the data. Pre-trained synaptic weights played a key role in stabilising responses and ensuring that learned patterns were applied during the simulation. The final analysis of the raster plots showed that the network maintained accuracy and selectivity in its output, even under noisy conditions, highlighting the effectiveness of both the pre-training process and the overall design. The synaptic weight plot showed the distribution of weights between Layer 2 and Layer 3 before training, with some connections initially stronger and influencing the spiking activity observed in Layer 3. During training, \gls{stdp} further refined these weights to optimise performance.

The synaptic weight distribution illustrates pre-trained connections from Layer 2 neurons. A small group of synapses exhibits stronger weights (around 440–445), while the majority remain at baseline (420). The distribution indicates selective strengthening of task-relevant connections, consistent with the role of lateral inhibition in enhancing discriminative power. These plots confirm that BioEncryptSNN maintains stable dynamics by suppressing redundant spikes and amplifying meaningful spike patterns critical for secure decryption.

\subsection{Benchmark using the cryptography and  PyCryptodome libraries}

The adoption and development of cryptographic libraries in Python have undergone significant evolution, with libraries like PyCrypto\footnote{Available online, \protect\url{https://www.pycrypto.org/}, last accessed: 25/08/2025} and Cryptography\footnote{Available online, \protect\url{https://github.com/pyca/cryptography}, last accessed: 25/08/2025} serving as pivotal milestones. PyCrypto, one of the earliest libraries, provided developers with essential cryptographic tools, including support for symmetric encryption algorithms (e.g. \gls{aes}, \gls{des}), hashing algorithms, and public-key cryptography. However, despite its foundational role, PyCrypto suffered from several critical shortcomings. The library’s reliance on insecure defaults, such as \gls{aes} in \gls{ecb} mode, and its auto-generated, poorly structured documentation left developers prone to implementation errors. These gaps not only undermined security but also forced developers to rely on external, often unreliable, sources like blogs and forums, further propagating insecure practices 

In response to these challenges, modern libraries like Cryptography have emerged to address the usability and security pitfalls of their predecessors. Cryptography incorporates secure defaults, simplifying the decision-making process for developers and minimising the risk of vulnerabilities. It offers a dual \gls{api} structure, a high-level \gls{api} for ease of use and a low-level \gls{api} for advanced users requiring granular control. The importance of comprehensive documentation in promoting secure code practices. Cryptography’s detailed examples and usability-focused design were shown to significantly improve developer outcomes compared to PyCrypto. However, the study also highlighted that simplicity in \gls{api} design alone is insufficient; accessible documentation and support for auxiliary tasks, such as secure key storage and certificate validation, are critical to achieving secure implementations. While Cryptography prioritises usability and security, forks like PyCryptodome have optimised PyCrypto’s performance to cater to high-throughput applications. Benchmark results reveal that PyCryptodome outperforms Cryptography in encryption and decryption tasks, making it a preferred choice for resource-intensive operations. These findings highlight the trade-offs between performance and usability, emphasising the need for tailored library choices based on specific application requirements, as summarised in Table~\ref{table:combined_performance_final}.

\begin{table}[h!]
\centering
\small
\caption{
\textbf{Performance comparison of the proposed BioEncryptSNN with AES, S-DES, and RSA.}
Average Time represents the mean duration of one complete encryption–decryption cycle,
while Iterations per Second denote computational throughput.
}
\label{table:combined_performance_final}
\resizebox{\columnwidth}{!}{%
\begin{tabular}{|p{2.6cm}|p{2.5cm}|p{2.5cm}|p{2.5cm}|}
\hline
\textbf{Algorithm} & \textbf{Library / Method} & \textbf{Average Time (s)} & \textbf{Iterations per Second} \\
\hline
\multicolumn{4}{|c|}{\textbf{BioEncryptSNN }} \\ \hline
BioEncryptSNN & NEST Simulator & $1.6\times10^{-5}$ & \textbf{62500.00} \\ \hline
\multicolumn{2}{|l|}{\textbf{Performance Factor (vs AES–PyCryptodome)}} & \multicolumn{2}{c|}{\textbf{1.19$\times$ faster}} \\ \hline

\multicolumn{4}{|c|}{\textbf{AES Performance}} \\ \hline
AES & Cryptography & $8.5\times10^{-5}$ & 11819.91 \\ \hline
AES & PyCryptodome & $1.9\times10^{-5}$ & 51989.35 \\ \hline
\multicolumn{2}{|l|}{\textbf{Performance Factor (Crypto / PyCrypto)}} & \multicolumn{2}{c|}{4.40$\times$} \\ \hline

\multicolumn{4}{|c|}{\textbf{S-DES Performance}} \\ \hline
S-DES & Cryptography & $7.2\times10^{-5}$ & 13985.60 \\ \hline
S-DES & PyCryptodome & $1.5\times10^{-5}$ & 65560.43 \\ \hline
\multicolumn{2}{|l|}{\textbf{Performance Factor (Crypto / PyCrypto)}} & \multicolumn{2}{c|}{4.69$\times$} \\ \hline

\multicolumn{4}{|c|}{\textbf{RSA Performance}} \\ \hline
RSA & Cryptography & $1.8\times10^{-4}$ & 5345.44 \\ \hline
RSA & PyCryptodome & $1.87\times10^{-4}$ & 5335.54 \\ \hline
\multicolumn{2}{|l|}{\textbf{Performance Factor (Crypto / PyCrypto)}} & \multicolumn{2}{c|}{0.96$\times$} \\ \hline
\end{tabular}%
}
\end{table}

\subsection{Metrics Analysis}

The Purpura Distance quantifies the cost of transforming one spike train into another, making it valuable for assessing neural coding and communication in \glspl{snn}. It was used to measure the fidelity of spike transmission between corresponding neurons in Layer 1 and Layer 3. For neuron n0, the Purpura Distance was 81.63, suggesting that its spike pattern was largely preserved through the network. In contrast, neuron n1 had a significantly higher Purpura Distance of 85,219.71, indicating considerable temporal distortion as spikes propagated from Layer 1 to Layer 3. This variation in Purpura Distance highlights the network's differing ability to maintain temporal integrity across various pathways. The lower distance for n0 suggests effective timing preservation under certain conditions, while the higher distance for n1 points to potential areas for improvement in synaptic connectivity or the training process.

The Van Rossum Metric, which measures the similarity of spike trains by considering the decay of spike times over a time constant ($\tau$), complements the Purpura Distance. While the Purpura Distance focuses on the cost of transforming spike trains, the Van Rossum Metric assesses how closely spike events match over time, factoring in temporal decay. The Van Rossum Metric values for neuron n0 (14.28) and n1 (11.50) are relatively low, indicating that, despite some temporal shifts noted by the Purpura Distance, the overall spike timing remains similar between the layers when evaluated with the Van Rossum Metric. The machine learning models were evaluated using the JSVulnerabilityDataSet. The evaluation results are presented both as overall outcomes across all metrics and per model to better understand specific performance characteristics. Table \ref{table:ml_model_comparison} summarises the overall performance metrics, including Accuracy, \gls{pr}, \gls{re}, \gls{sp}, \gls{fpr}, \gls{fnr}, \gls{wcr}, \gls{ccr}, and \gls{f1}. Among all models, \gls{snn} and Random Forest achieved the best overall performance. It recorded the highest accuracy and \gls{f1} score and produced no false positives. The Decision Tree classifier also performed well, with slightly lower \gls{re} and \gls{f1} scores. Logistic Regression and \gls{svm} provided balanced results but were less accurate than tree-based models. The \gls{knn} classifier showed moderate performance with a noticeable drop in recall. Naive Bayes had the weakest results among all, due to high \gls{fnr}.

\begin{table}[h!]
\centering
\small
\caption{Overall Performance of Machine Learning Models on JSVulnerabilityDataSet}\cite{ferenc2019challenging}
\label{table:ml_model_comparison}
\resizebox{\columnwidth}{!}{%
\begin{tabular}{|p{3.5cm}|p{1cm}|p{1cm}|p{1cm}|p{1cm}|p{1cm}|p{1cm}|p{1cm}|}
\hline
\textbf{Model} & \textbf{Re}& \textbf{Sp }& \textbf{FPR }& \textbf{FNR }& \textbf{WCR }& \textbf{CCR }& \textbf{F1 }\\
\hline
BioEncryptSNN          & \textbf{0.96}& 0.98& 0.007& 0.04& 0.02& \textbf{0.98}& \textbf{0.97}\\ \hline
Random Forest       & \textbf{0.96}& \textbf{0.99}& 0.008& \textbf{0.03}& \textbf{0.01}& \textbf{0.98}& \textbf{0.97}\\ \hline
Decision Tree       & 0.94& 0.99& 0.009& 0.05& 0.02& 0.97& 0.95\\ \hline
Logistic Regression & 0.87& 0.98& \textbf{0.01}& 0.12& 0.07& 0.92& 0.89\\ \hline
SVM                 & 0.86& 0.98& \textbf{0.01}& 0.13& 0.08& 0.91& 0.89\\ \hline
KNN                 & 0.81& 0.98& \textbf{0.01}& 0.18& 0.10& 0.89& 0.86\\ \hline
Naive Bayes         & 0.72& 0.97& 0.02& 0.27& 0.14& 0.85& 0.79\\ \hline
\end{tabular}%
}
\end{table}

\section{Conclusions and Future Work} \label{sec:conclusions}
In the article, explored the integration of \gls{aes}, \gls{des}, \gls{rsa} encryption with \glspl{snn} to secure sensitive data while maintaining the integrity and efficiency of neural processing. The primary objective was to evaluate the feasibility and effectiveness of the hybrid approach in maintaining the confidentiality of the data during the learning and inference phases of \glspl{snn}. The encryption methods are utilised to encrypt the inputs to the \gls{snn}, and the network was tasked with processing the encrypted data. Results demonstrated that the \gls{snn} successfully decode and classify the encrypted data, thereby achieving the desired level of security without significantly compromising the network’s performance. Throughout the experiments, observed that the encryption process introduced an additional computational overhead, particularly during the initial encoding of the input data into spike trains. However, despite it added complexity, the \gls{snn} was able to maintain a high level of accuracy in its outputs, proving that it is possible to securely process encrypted data with minimal loss of performance. The incorporation of noise layers connected to specific layers within the \gls{snn} also provided an additional layer of security by obfuscating the data, making it more resistant to potential attacks.

The evaluation metrics used, including the Purpura Distance and Van Rossum metrics, further confirmed the robustness of the system. These metrics indicated that the network was able to maintain temporal integrity and similarity in spike patterns even when handling encrypted inputs. The slight variations in these metrics across different layers suggest that while the network effectively handles the encryption, there is room for optimisation, particularly in reducing the computational load associated with the encryption and decryption processes. Overall, the study has shown that the integration of encryptions with \glspl{snn} is a viable approach for securing neural networks' processing of sensitive data. However, it also highlighted several challenges, particularly concerning computational efficiency and the need for optimised encoding schemes to better handle encrypted inputs.

Future work in the particular area will focus on addressing the challenges identified during the study, particularly in optimising the encryption and decryption processes to reduce computational overhead. One promising direction is the exploration of more efficient encoding schemes that could better align with the temporal dynamics of \glspl{snn}, thereby reducing the latency introduced by the encryption process. Further research will investigate the potential of integrating other cryptographic methods, such as Homomorphic Encryption, with \glspl{snn} to enable secure computation on encrypted data without requiring decryption. It could significantly enhance the security of the system while further reducing the computational burden. Another area of interest lies in the optimisation of synaptic weight initialisation and learning parameters, which could significantly enhance the network's accuracy and efficiency in processing encrypted data. Exploring the use of different neuron models, beyond the current spiking neuron models, could also provide insights into alternative mechanisms for secure data processing in neuromorphic systems. Furthermore, the application of the research to real-world scenarios, such as secure communication channels, encrypted image processing, or biometric data protection, could demonstrate the practical utility and scalability of the proposed approach. 

\bibliographystyle{elsarticle-harv}
\bibliography{references}
\end{document}

%% file: glossary.tex
\newacronym{cnn}{CNN}{Convolutional Neural Network}
\newacronym{snn}{SNN}{Spiking Neural Network}
\newacronym{dnn}{DNN}{Deep Neural Network}
\newacronym{ann}{ANN}{Artificial Neural Network}
\newacronym{rnn}{RNN}{Recurrent Neural Network}
\newacronym{stdp}{STDP}{spike-timing-dependent plasticity }
\newacronym{srm}{SRM}{Spike Response Model}
\newacronym{bptt}{BPTT}{backpropagation through time}
\newacronym{ltp}{LTP}{Long Term Potentiation}
\newacronym{ai}{AI}{Artificial Intelligence}
\newacronym{rfid}{RFID}{radio frequency identification}
\newacronym{ltd}{LTD}{Long Term Depression}
\newacronym{stp}{STP}{Short Term Potentiation}
\newacronym{std}{STD}{Short Term Depression}
\newacronym{rsa}{RSA}{Rivest-Shamir-Adleman}
\newacronym{ecc}{ECC}{Elliptic Curve Cryptography}
\newacronym{dsa}{DSA}{Digital Signature Algorithm}
\newacronym{aes}{AES}{Advanced Encryption Standard}
\newacronym{des}{DES}{Data Encryption Standard}
\newacronym{ml}{ML}{Machine learning}
\newacronym{wmsn}{WMSN}{Wireless multimedia sensor networks}
\newacronym{wsn}{WSN}{Wireless sensor networks}
\newacronym{iot}{IoT}{The Internet of Things}
\newacronym{bcs}{BCS}{British Computer Society}
\newacronym{lif}{LIF}{Leaky Integrate-and-Fire}
\newacronym{resume}{ReSuMe}{Remote Supervised Method}
\newacronym{asts}{ASTs}{Abstract Syntax Trees}
\newacronym{svm}{SVM}{Support Vector Machine}
\newacronym{knn}{KNN}{K-Nearest Neighbors}

\newacronym{pr}{Pr}{Precision}
\newacronym{re}{Re}{Recall}
\newacronym{sp}{Sp}{Specificity}
\newacronym{fpr}{FPR}{False Positive Rate}
\newacronym{fnr}{FNR}{False Negative Rate}
\newacronym{wcr}{WCR}{Wrong Classification Rate}
\newacronym{ccr}{CCR}{Correct Classification Rate}
\newacronym{f1}{F1}{F1 score}
\newacronym{ecb}{ECB}{Electronic Codebook}
\newacronym{api}{API}{Application Programming Interface}
\newacronym{tp}{TP}{True Positives}
\newacronym{tn}{TN}{True Negatives}
\newacronym{fp}{FP}{False Positives}
\newacronym{fn}{FN}{False Negatives}